\documentclass{article}
\usepackage[table,xcdraw,dvipsnames]{xcolor}
\usepackage{conference, times}
\usepackage{hyperref}
\usepackage{url}
\usepackage[frozencache,cachedir=./cache]{minted}
\usepackage{booktabs}
\usepackage{multirow}
\usepackage[many, most, skins]{tcolorbox}
\usepackage{pifont}
\usepackage{fontawesome5}
\usepackage[ruled,linesnumbered]{algorithm2e}
\usepackage{xspace}
\usepackage{ifthen}
\usepackage{enumitem}
\usepackage{tikz}
\usepackage[normalem]{ulem}
\usepackage{tabularx}
\usepackage{enumitem}
\usepackage{tabularx}
\usepackage{subcaption}
\usepackage{arydshln}
\usepackage{array}

\tcbuselibrary{listings}
\tcbuselibrary{minted}
\usetikzlibrary{arrows, calc, shapes.geometric, patterns}

\setlist[itemize]{nosep}
\setlist[enumerate]{nosep}

\newcolumntype{C}[1]{>{\centering\arraybackslash}p{#1}}
\newcolumntype{M}[1]{>{\centering\arraybackslash}m{#1}}

\newcommand\circnum[1]{\ding{\numexpr181+#1\relax}}
\newcommand\circnumwhite[1]{\ding{\numexpr191+#1\relax}}
\newcommand\figref[1]{Figure~#1\xspace}
\newcommand\tabref[1]{Table~#1\xspace}

\makeatletter
\newcommand{\namelabel}[1]{%
  \phantomsection
  \renewcommand{\@currentlabel}{#1}%
  \label{#1}%
}
\makeatother

\newcounter{findingcounter}
\newcommand{\finding}[1]{
\begin{tcolorbox}[
    left=4pt, right=4pt, top=2pt, bottom=2pt,
    boxrule=0.2mm,
    leftrule=1mm,
    arc=0mm,
    colframe=black!40!white,
    colback=white,
    colbacktitle=black!50!white
]
\refstepcounter{findingcounter}
\textbf{Finding \thefindingcounter\namelabel{\thefindingcounter}:~}{#1}
\end{tcolorbox}
}

\newcolumntype{\CeX}{>{\centering\let\newline\\\arraybackslash}X}%
\newcommand{\TwoSymbolsAndText}[3]{%
  \renewcommand{\arraystretch}{0}
  \begin{tabularx}{\textwidth}{c\CeX c}%
    #1 & #2 & #3
  \end{tabularx}%
}
\newtcblisting[use counter=metaprompt, number format=\arabic]{metaprompt}[2]{
  listing engine=minted,
  minted language=text,
  minted options={autogobble,breaklines,breaksymbolleft={},breaksymbolright={},escapeinside=||,fontsize=\small},
  listing only,
  size=title,
  arc=1.5mm,
  breakable,
  enhanced jigsaw,
  colframe=gray,
  coltitle=white,
  boxrule=0.4mm,
  colback=white,
  coltext=black,
  title=\TwoSymbolsAndText{\faCode}{%
    \textbf{Prompt \thetcbcounter}\ifthenelse{\equal{#1}{}}{}{\textbf{:} \textbf{\textit{#1}}}%
  }{\faCode},
  label=metaprompt:#2
}

\newcommand\dashline{\arrayrulecolor{gray}\hdashline[1pt/1pt]\arrayrulecolor{black}}
\newcommand\pdashline[1]{\arrayrulecolor{gray}\cdashline{#1}[1pt/1pt]\arrayrulecolor{black}}

\iclrfinalcopy
\begin{document}

\title{Re-Evaluating Code LLM Benchmarks Under Semantic Mutation}

\author{Zhiyuan Pan, Xing Hu, Xin Xia, Xiaohu Yang \\
  Zhejiang University \\
  Hangzhou, China \\
  \texttt{\{zy\_pan, xinghu\}@zju.edu.cn, xin.xia@acm.org, yangxh@zju.edu.cn}
}

\maketitle

\begin{abstract}

In the era of large language models (LLMs), code benchmarks have become an important research area in software engineering and are widely used by practitioners.
These benchmarks evaluate the performance of LLMs on specific code-related tasks, such as code understanding and generation.
A critical step in constructing code benchmarks is the design of prompts. However, as existing code benchmarks typically rely on a single prompt template per task, they are prone to the issue of \textbf{prompt sensitivity}, where minor prompt variations could result in substantial performance variations, leading to unreliable evaluations of model capabilities.

While previous studies have explored prompt sensitivity, their experimental designs and findings are limited to traditional natural language processing (NLP) tasks. 
In this paper, we present an empirical study to investigate prompt sensitivity in code benchmarks. We first propose a general framework that modifies prompt templates in a manner that preserves both their semantics and their structure as much as possible. Based on the framework, we conduct extensive experiments across eight code benchmark tasks on 10 representative open-source LLMs, with each task featuring 100 semantically similar prompt templates. We then analyze the evaluation results using various statistical metrics, focusing on both absolute and relative model performance. Our findings suggest that even slight prompt variations can lead to significant shifts in performance. Additionally, we observe that such variations can introduce inconsistencies in the performance rankings across different models. These insights highlight the need for considering prompt sensitivity when designing future code benchmarks, to ensure more reliable and accurate evaluation of LLM capabilities.

\end{abstract}

\section{Introduction}
\label{sec:intro}

Large language models (LLMs)~\cite{openai2024gpt4ocard,llama3,bai2023qwentechnicalreport,codellama} have been widely applied in software engineering (SE)-related fields, and they have shown remarkable performance in various code-related tasks, such as code generation~\cite{humaneval}, test case generation~\cite{tufano2020unit} and program understanding~\cite{lu2021codexgluemachinelearningbenchmark, chen2024reasoningruntimebehaviorprogram}.

To evaluate the capabilities of LLMs on specific code-related tasks, many code benchmarks~\cite{humaneval,mbpp,du2023classeval,testeval,cruxeval,chen2024reasoningruntimebehaviorprogram} have been proposed.
These benchmarks are crucial for SE practitioners, as researchers rely on their results to identify areas for improvement, while end users use them to select the most capable model for their specific use cases.
The creation and use of code benchmarks typically involves the following parts~\cite{cao2025buildbenchmarkrevisiting274}: preparing datasets, designing prompts, interacting with LLMs, and performing tests. 
A critical step in this process is designing prompts, which generally includes defining task requirements and providing reference information~\cite{google_vertex_ai_prompt_design}. 
However, since prompt design is inherently subjective, the choice of prompts can influence the evaluation results for LLMs~\cite{amatriain2024promptdesignengineeringintroduction}.
Therefore, we argue that a well-designed code benchmark should be \textbf{robust}: it should accurately estimate the knowledge boundaries~\cite{knowledge_boundary} of an LLM and provide consistent performance rankings across a range of semantically similar prompt choices.

\begin{figure}[t]
    \centering
    \includegraphics[width=0.6\linewidth]{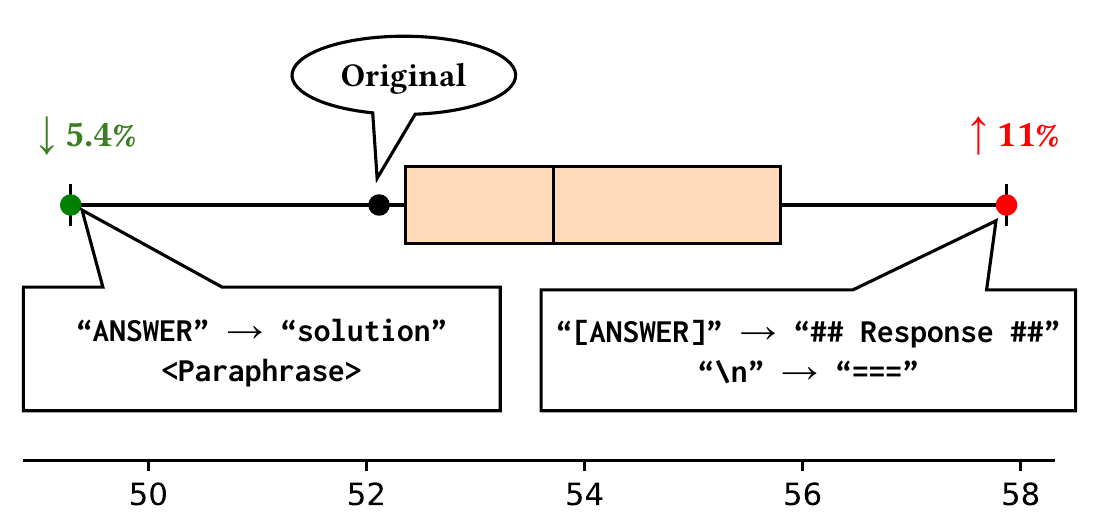}
    \caption{Distribution of evaluation results for CRUXEval~\cite{cruxeval} with 100 semantically similar prompt templates on CodeLlama-7B-Instruct~\cite{codellama}. Slight mutations (presented in the rectangles) lead to significant variations in model performance (pass@5 score), with changes ranging from a 5.4\% decrease to an 11\% increase.}
    \label{fig:intro}
\end{figure}

However, we find that minor changes to prompts in code benchmarks can lead to substantial variations in model performance, a phenomenon referred to as \textbf{prompt sensitivity}. 
For example, the boxplot in \figref{\ref{fig:intro}} illustrates the distribution of evaluation results for a code reasoning task in a benchmark named CRUXEval~\cite{cruxeval}, on the \textit{CodeLlama-7B-Instruct}~\cite{codellama} model using 100 mutated prompt templates. 
The results show significant variation in model performance, with a maximum decrease of 5.4\% and a maximum increase of 11\% compared to the original metric value. 
Despite these substantial performance differences, the modifications to the prompts are subtle and semantically similar, suggesting that prompt sensitivity is likely common across other code benchmarks.
Prompt sensitivity introduces limitations in current code benchmarks: with only one prompt template for evaluation, (1) the absolute performance of an individual LLM may not be accurately estimated, and (2) the relative performance across multiple LLMs may be skewed, leading to unreliable model rankings.

This preliminary finding aligns with previous studies on prompt sensitivity. Sclar et al.~\cite{sclar2023quantifying}, Mizrahi et al.~\cite{mizrahi-etal-2024-state}, and Cao et al.~\cite{cao_nips2024} investigate prompt sensitivity by exploring semantically similar mutations of prompts, including modifications to punctuation and spacing, and whole-prompt paraphrasing.
However, these studies focus on traditional NLP benchmarks~\cite{efrat-etal-2023-lmentry, cao_nips2024, wang-etal-2022-super, Srivastava2023BeyondTI, suzgun-etal-2023-challenging}, with code benchmarks remaining unexplored. Compared to these NLP benchmarks used in previous studies, model responses and the evaluation process in code benchmarks differ in several ways: (1) input and output sequences are typically longer, (2) prompts are often more structural due to the common inclusion of complex referential contexts and few-shot examples, and (3) while NLP benchmarks commonly use text similarity-based measures (e.g., exact match) to check for correctness, many recent code benchmarks rely on execution-based measures (e.g., test cases). As these differences are related to the assessment of prompt sensitivity, it may not be appropriate to directly extend conclusions from previous studies to code benchmarks without further experimentation.
Additionally, given the more complex and structural nature of prompts in code benchmarks, the designs of existing studies may be insufficient for thoroughly exploring the prompt space in these benchmarks.
These factors underscore the need for an extensive study of prompt sensitivity in recent code benchmarks.

In this paper, we present an empirical study on prompt sensitivity in code benchmarks. To enable an extensive study, we select eight tasks from three code benchmarks, prepare 100 semantically similar prompt templates for each task, and conduct experiments on ten representative LLMs. Our approach to mutating prompt templates consists of two components:

\circnum{1} \textbf{Offline step.} For each task, we first manually decompose the original prompt template into a syntax tree. We then design a variety of operations that modify the tree nodes to explore various subtle mutations.

\circnum{2} \textbf{Online step.} Using an LLM, we generate diverse input arguments for the operations defined in the offline step. To create a large number of mutations, we iteratively generate input arguments and apply the operations, starting from the original prompt template.

Throughout these steps, we ensure that mutations do not compromise the original semantics or structure of the prompt by restricting the types of operation and employing a validation process. After generating the required number of mutated prompt templates, we perform the standard evaluation steps in the code benchmarks and collect the results for analysis.
In particular, we investigate the following research questions:

\textbf{RQ1 (Impact on absolute model performance):} \textit{How do semantically similar modifications to prompt templates affect the performance of individual LLMs in code-related tasks?}

\textbf{RQ2 (Impact on relative model performance):} \textit{How do semantically similar modifications to prompt templates affect the rankings of multiple LLMs in code-related tasks?}

With these research questions, we investigate whether prompt sensitivity commonly exists for the selected code benchmark tasks and LLMs, and whether it affects the robustness of the benchmark tasks. Based on our results, we have the following findings. In RQ1, we find that slight prompt variations in our experiments can cause substantial variations in model performance, suggesting that prompt sensitivity cannot be neglected in code benchmarks. Given the performance variations, substantial performance gain can also be achieved in these tasks simply with our mutated prompt templates. However, for a specific benchmark task, prompt sensitivity does not have a clear correlation with model capability. Thus, a more capable model does not necessarily indicate more stable responses. In RQ2, we find that prompt variations can introduce inconsistencies to model rankings, especially to rankings within a particular model family (e.g., models in the Llama family). Moreover, the top-performing prompt templates for specific code-related tasks exhibit poor overlap across different models. Thus, current code benchmarks may not provide reliable capability rankings across models. Our findings highlight the need for more robust code benchmarks that account for prompt sensitivity by incorporating multiple prompt templates.

\noindent\textbf{Contributions.}
In summary, we make the following contributions:

\begin{enumerate}
    \item To the best of our knowledge, we are the first to provide a comprehensive analysis of the impact of prompt sensitivity in the field of software engineering.
    \item We introduce a general framework for prompt template mutation in code benchmarks, designed to support our study.
    \item We conduct extensive experiments on recent code benchmarks and large language models (LLMs) to investigate prompt sensitivity in code-related tasks, providing practical findings and implications for the software engineering community.
\end{enumerate}

\noindent\textbf{Paper Organization.}
The remainder of the paper is structured as follows. Section 2 introduces the background of our study and summarizes related work. Section 3 presents the design of our study. Section 4 describes the implementation of our study, including evaluation targets and experimental settings. Section 5 presents the experimental results and our findings based on the results. Section 6 discusses the results of case studies and threats to validity. Section 7 concludes the paper.

\section{Background and Related Work}

\subsection{LLMs for Software Engineering}
To obtain better performance in the software engineering field, large language models for code (i.e. code LLMs)~\cite{codellama, codegemmateam2024codegemmaopencodemodels, hui2024qwen25codertechnicalreport, guo2024deepseekcoderlargelanguagemodel} have been proposed. Compared with their corresponding foundation models, code LLMs are further fine-tuned on code datasets. Previous work has shown the promising performance of LLMs in various code-related tasks, such as code understanding~\cite{lu2021codexgluemachinelearningbenchmark}, code generation~\cite{humaneval}, test case generation~\cite{tufano2020unit}, and program repair~\cite{program_repair}. 

To evaluate the performance of LLMs on code-related tasks, many code benchmarks have been proposed. 
To evaluate code-writing capabilities, HumanEval~\cite{humaneval} and MBPP~\cite{mbpp} are two representative benchmarks. To improve practicality, some subsequent work~\cite{du2023classeval, liu2023repobench, yu2024codereval} extends the evaluation to the more challenging context-aware scenarios, while other work~\cite{jain2024livecodebenchholisticcontaminationfree} addresses the potential data leakage~\cite{zhou2025lessleakbenchinvestigationdataleakage} issue.
In addition, test case generation is also an important field of research. For this task, Wang et al.~\cite{testeval} challenge LLMs to write test cases for competition-level programs (e.g., solution programs in LeetCode), while Zeng et al.~\cite{coderujb} extends the evaluation to the more complex project-level scenarios.
Apart from classic coding tasks, there are also benchmarks that evaluate the capabilities of LLMs from novel perspectives. 
Lu et al. propose CodeXGLUE~\cite{lu2021codexgluemachinelearningbenchmark}, which contains various tasks to evaluate code understanding.
Additionally, Gu et al.~\cite{cruxeval} and Chen et al.~\cite{chen2024reasoningruntimebehaviorprogram} propose to evaluate code reasoning abilities using various properties of Python programs.

\subsection{Prompt Sensitivity}

Designing prompts is crucial for model users and benchmark creators. Prompt sensitivity refers to the robustness of LLMs in producing consistent responses when prompts are slightly modified, and is usually evaluated using model performance variations.

Existing work has explored different ways to mutate prompts to study the impact of prompt sensitivity. To study the impact of prompt sensitivity on the \textit{safety of LLM generations}, a line of research focuses on mutating prompts in ways that alter their original semantic meanings, such as modifying existing information or inserting additional information. Zhu et al.~\cite{promptbench} propose an evaluation framework named PromptBench to construct adversarial attacks through four types of prompt mutation: character-, word-, sentence- and semantic-level. Salinas and Morstatter~\cite{salinas-morstatter-2024-butterfly} evaluate the effect of prompt variations with strategies such as jailbreaking (i.e., using crafted inputs to bypass ethical or legal constraints on LLMs). 

In contrast, another line of research focuses on the impact on \textit{robust model evaluation}, and explores semantically similar prompt mutations. Sclar et al.~\cite{sclar2023quantifying} evaluate the influence of prompt choice by exploring different prompt formats, such as separators and spacing. Mizrahi et al.~\cite{mizrahi-etal-2024-state} propose to paraphrase the entire prompts with various strategies and manually check for validity. While these studies conduct experiments at task-level granularity (i.e., modify the general prompt template of a task and fill the template with data samples for evaluation), Cao et al.~\cite{cao_nips2024} and Zhuo et al.~\cite{zhuo-etal-2024-prosa} propose to paraphrase and evaluate prompts at case-level granularity (i.e., modify the specific prompt for each data sample in a task). 

In this paper, we study whether current code benchmarks reliably evaluate absolute and relative model capabilities. As introduced in Section~\ref{sec:intro}, compared to the NLP benchmarks studied in previous work, the differences and complexities of code benchmarks motivate our empirical study.
In particular, we follow the latter line of research, i.e., semantically similar prompt mutations. Compared to previous studies that focus exclusively on format modification or simple whole-text paraphrasing, we propose an experimental approach that integrates a broader range of fine-grained mutations. Additionally, we maintain the original prompt structures during the mutation process.

\section{Study Design}
\label{sec:design}

In this section, we first present an overview of the methodology of our study. Then we describe the components in detail.

\subsection{Overview}
\label{sec:overview}

\tikzset{
    >=stealth',
    punkt/.style={
           rectangle,
           rounded corners,
           draw=black, thick,
           text width=5em,
           minimum height=2em,
           },
    pil/.style={
           ->,
           thick,
           shorten <=1pt,
           shorten >=1pt,
           },
    alabel/.style={
           above,
           align=center,
           font=\small,
           },
    blabel/.style={
           below,
           align=center,
           font=\small,
           },
    llabel/.style={
           left,
           align=center,
           font=\small,
           },
    rlabel/.style={
           right,
           align=center,
           font=\small,
           },
    container/.style={
           rectangle,
           rounded corners,
           draw=black, thick,
           align=left,
           inner sep=1ex,
           font=\small,
           },
}

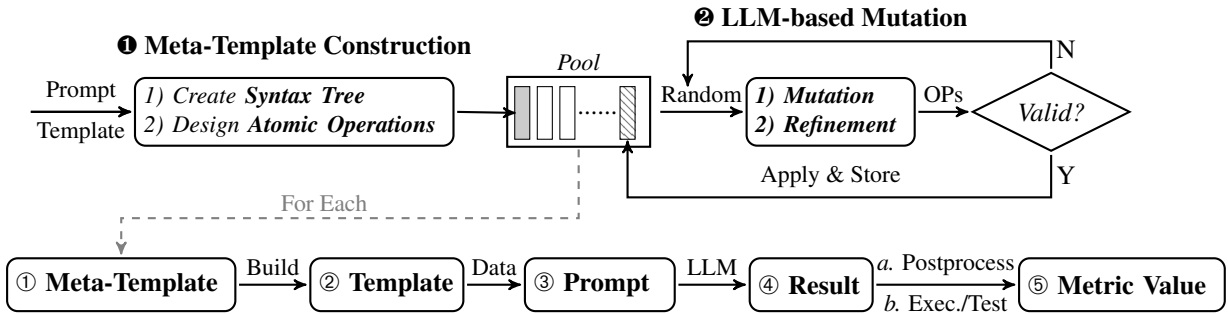
\begin{figure*}[htbp]%
\begin{tikzpicture}
\centering
 \node[punkt, text width=4cm] (o1) at (0,2) {
     \textit{\small{1) Create \textbf{Syntax Tree} \\ 2) Design \textbf{Atomic Operations}
    }}
 };
 \node at ([yshift=0.4cm]o1.north) {\circnum{1}~\textbf{Meta-Template Construction}};
 
\coordinate (SW) at ([xshift=0.7cm]o1.south east);
\coordinate (NE) at ([xshift=2.6cm]o1.north east);
\draw[thick] (SW) rectangle (NE);
\draw[fill=gray!40] ($(SW) + (0.1, 0.1)$) rectangle ($(SW) + (0.3, 0.8)$);
\draw ($(SW) + (0.4, 0.1)$) rectangle ($(SW) + (0.6, 0.8)$);
\draw ($(SW) + (0.7, 0.1)$) rectangle ($(SW) + (0.9, 0.8)$);
\node at ($(SW) + (1.2, 0.45)$) {......};
\draw[pattern=north west lines,pattern color=gray] ($(SW) + (1.5, 0.1)$) rectangle ($(SW) + (1.7, 0.8)$);
\node at ([xshift=1.65cm, yshift=0.2cm]o1.north east) {\textit{\small{Pool}}};

 \node[punkt, text width=2cm] (o2) at ([xshift=5cm]o1.east) {
    \textbf{\textit{{\small{1) Mutation \\ 2) Refinement}}}}
 };
 \node at ([yshift=0.8cm]o2.north) {\circnum{2}~\textbf{LLM-based Mutation}};

 \node[draw, thick, diamond, aspect=2] (o3) at ([xshift=1.8cm]o2.east) {
    \textit{Valid?}
 };
 \node at ([xshift=0.2cm, yshift=-0.3cm]o3.south) {Y};
 \node at ([yshift=-0.4cm]o2.south) {\small{Apply \& Store}};
 \node at ([xshift=0.2cm, yshift=0.3cm]o3.north) {N};

\draw[pil] ([xshift=-40pt]o1.west) -- node[alabel] {Prompt} (o1.west);
\draw[pil] ([xshift=-40pt]o1.west) -- node[blabel] {Template} (o1.west);
\draw[pil] (o1.east) -- ($(SW) + (0.15, 0.45)$);
\draw[pil] ([xshift=2.7cm]o1.east) -- node[alabel] {Random} (o2.west);
\draw[pil] (o2.east) -- node[alabel] {OPs} (o3.west);
\draw[pil] (o3.south) --++ (0,-0.6) -| ($(SW) + (1.6, 0.12)$);
\draw[pil] (o3.north) --++ (0,0.4) -| ([xshift=3.1cm,yshift=0.3cm]o1.east);
 
 \node[punkt, text width=2.8cm] (n1) at (-2.3, -0.3) {
    \circnumwhite{1}~\textbf{Meta-Template}
 };

 \node[punkt, text width=1.8cm] (n2) at ([xshift=2cm]n1.east) {
    \circnumwhite{2}~\textbf{Template}
 };

 \node[punkt] (n3) at ([xshift=1.8cm]n2.east) {
    \circnumwhite{3}~\textbf{Prompt}
 };

 \node[punkt, text width=1.4cm] (n4) at ([xshift=1.8cm]n3.east) {
    \circnumwhite{4}~\textbf{Result}
 };

 \node[punkt, text width=2.5cm] (n5) at ([xshift=3.3cm]n4.east) {
    \circnumwhite{5}~\textbf{Metric Value}
 };

 \node[gray] at ([xshift=0.4cm,yshift=-0.75cm]o1.south) {\small{For Each}};

\draw[pil, gray, dashed] ([xshift=1.65cm]o1.south east) --++ (0,-0.95) -| (n1.north);
\draw[pil] (n1.east) -- node[alabel] {Build} (n2.west);
\draw[pil] (n2.east) -- node[alabel] {Data} (n3.west);
\draw[pil] (n3.east) -- node[alabel] {LLM} (n4.west);
\draw[pil] (n4.east) -- node[alabel] {\textit{a.} Postprocess} node[blabel] {\textit{b.} Exec./Test} (n5.west);

\end{tikzpicture}
\caption{Overview of our study design. \circnum{1} and \circnum{2} serve as the data preparation phase. \circnumwhite{1} to \circnumwhite{5} represent the execution steps.}
\label{fig:approach}
\end{figure*}

For the sake of clarity, we first provide definitions of key terms used in our methodology:

\begin{itemize}
    \item \textbf{Prompt Template}: A template string for generating prompts. In our study, a benchmark task corresponds to a prompt template. A prompt is instantiated by filling the prompt template with benchmark data.
    \item \textbf{Syntax Tree}: A tree structure representing the hierarchical structure of a prompt template, constructed by parsing the prompt template and identifying its structural components. 
    Conversely, a prompt template can be built by traversing its syntax tree.
    \item \textbf{Atomic Operation}: A basic operation that modifies a single node of a syntax tree to create a new syntax tree.
    \item \textbf{Meta-Template}: A higher-level template consisting of a syntax tree and its available atomic operations. Meta-templates are used to generate semantically similar prompt templates.
\end{itemize}

\figref{\ref{fig:approach}} illustrates the overall \textit{Mutation-Inference-Evaluation} workflow of our study. The core component is the mutation step, where we start from the original prompt template of each selected benchmark task and generate a specific number of new semantically similar templates via \textbf{Meta-Template Construction} and \textbf{LLM-based Mutation} (i.e., \circnum{1} and \circnum{2} in \figref{\ref{fig:approach}}). Then, for each template, we perform the common \textit{inference and evaluation} process (i.e., \circnumwhite{1} $\rightarrow$ \circnumwhite{5} in \figref{\ref{fig:approach}}). In the remaining part of this section, we describe in detail how we perform steps \circnum{1} and \circnum{2} for a benchmark task.

\subsection{Meta-Template Construction}
\label{sec:meta-template}

Due to the diversity of software engineering tasks, it is not practical to design a general pattern to guide mutations for all kinds of prompt templates.
Thus, we construct meta-templates with task-specific operations for modifications. 
Specifically, we \textit{manually} construct a meta-template from the original prompt template as the initial seed for mutation, based on the following rationales:

\begin{enumerate}[label=\textbf{R\theenumi}]
    \item Respect of the original prompt structure
    \item Respect of the original prompt semantics
\end{enumerate}

As introduced in Section~\ref{sec:overview}, a meta-template consists of a syntax tree and its available atomic operations. To follow \textbf{R1}, we propose to build syntax trees of prompt templates and \textbf{only modify existing nodes} instead of adding or deleting nodes. To follow \textbf{R2}, we propose to design \textbf{semantically similar operations} on tree nodes for a specific prompt template. After constructing the initial meta-template, we put it in an empty seed pool and proceed to the \textit{LLM-based Mutation} step (Section~\ref{sec:llm_mut}).

\subsubsection{Create Syntax Tree}

\begin{figure}[htbp]
    \centering
    \includegraphics[width=0.6\linewidth]{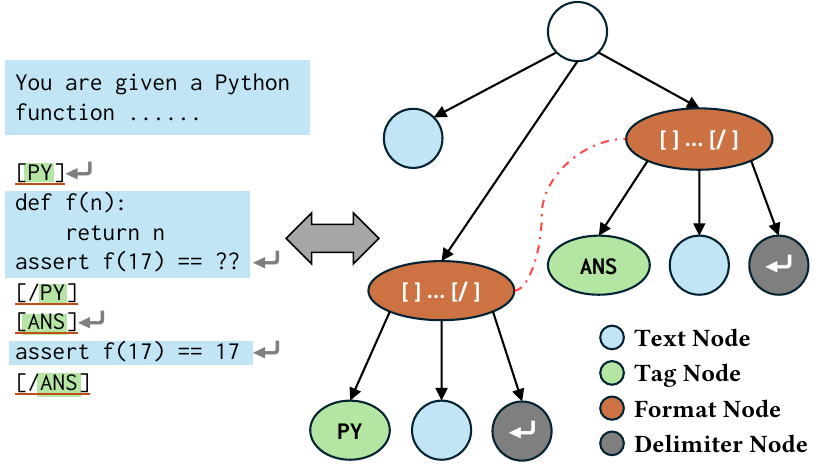}
    \caption{An example of a prompt template (partial; simplified for demonstration) in CRUXEval benchmark, and its corresponding syntax tree}
    \label{fig:syntax_tree}
\end{figure}

Compared with NLP benchmarks evaluated in previous work, the prompt templates of recent software engineering benchmarks exhibit more \textbf{complex structures}. In specific, to guide LLMs to fulfill the required tasks, these templates provide not only basic task requirements, but also structured \textit{sections} of additional information (i.e., task-specific code context). The structural prompt templates can be expressed by syntax trees.
To accurately capture these structures, we define four types of nodes:

\begin{itemize}
    \item \textbf{Text Node}: Represents textual information and code blocks.
    \item \textbf{Format Node}: Represents the format strings used for section headers, and for footers if they are present. In the example shown in \figref{\ref{fig:syntax_tree}}, the format node is represented as ``\texttt{[\{\}]...[\textbackslash\{\}]}'', where ``\texttt{[\{\}]}'' denotes the header, ``\texttt{[\textbackslash\{\}]}'' denotes the footer, and ``\texttt{\{\}}'' represents the text placeholder. For simplicity, we treat all duplicate format nodes within a syntax tree as a single shared instance, as illustrated by the red link in \figref{\ref{fig:syntax_tree}}.
    \item \textbf{Tag Node}: Represents the text in section headers, and in footers if they are present. In the example shown in \figref{\ref{fig:syntax_tree}}, the tag nodes are ``\texttt{PY}'' and ``\texttt{ANS}''. Inserting these tags into the format node produces complete section headers or footers.
    \item \textbf{Delimiter Node}: Represents boundaries between sections, such as line breaks (i.e., the delimiter in the example in \figref{\ref{fig:syntax_tree}}) or symbols like ``\texttt{\`{}\`{}\`{}}''.
\end{itemize}

Based on these types of nodes, we manually construct a syntax tree for a prompt template by parsing the text, assigning node types, and building the tree hierarchy.

\subsubsection{Design Atomic Operations}

To ensure diverse mutations and a thorough exploration of the prompt space, we define five types of atomic operations for different node types:

\begin{center}
\centering
\begin{tabularx}{0.6\columnwidth}{@{}>{\bfseries}l@{\hspace{3em}}X@{}}
\textbf{Node Type}  & \textbf{Operation Type} \\ \midrule
Text Node           &
\textbf{1.} Paraphrase textual information
\\ \dashline
\multirow{2}{*}{Tag Node}   & 
\textbf{2.} Paraphrase tag content
\\ 
                    &
\textbf{3.} Change tag casing
\\ \dashline
Format Node         & 
\textbf{4.} Change format string
\\ \dashline
Delimiter Node      & 
\textbf{5.} Change delimiter string
\end{tabularx}
\end{center}

\begin{figure}[t]
    \centering
    \includegraphics[width=0.6\linewidth]{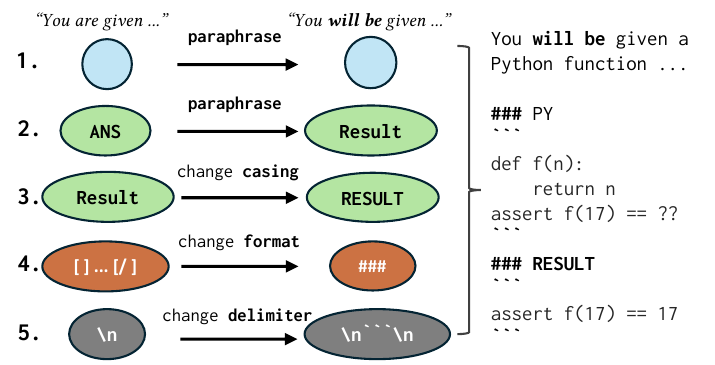}
    \caption{Examples of atomic operations and the resulting prompt template (corresponding to \figref{\ref{fig:syntax_tree}})}
    \label{fig:op_example}
\end{figure}

After creating the syntax tree, we \textit{manually} create detailed atomic operations, along with their description and required arguments, for specific nodes based on the previous definition. 
This step is specific to the benchmark task. For illustration, we use the prompt template and syntax tree shown in \figref{\ref{fig:syntax_tree}} and provide brief examples of each type of atomic operation in \figref{\ref{fig:op_example}}, along with the resulting prompt template obtained after executing the operations sequentially.

Regarding our goals \textbf{R1} (prompt structure) and \textbf{R2} (prompt semantics), the operations clearly adhere to \textbf{R1} since they neither add nor delete nodes.
For \textbf{R2}, the final mutation of the syntax tree can be decomposed into a sequence of atomic operations. While an individual operation is designed to maintain semantic similarity, a sequence of operations may unintentionally break the original semantics. We describe a rule-based refinement and validation process employed during mutation in Section~\ref{sec:refinement} and~\ref{sec:validation}. Additionally, after generating all mutated templates, we manually double-check to ensure that the semantics are preserved (Section~\ref{sec:manual_check}).

\subsection{LLM-based Mutation}
\label{sec:llm_mut}

\begin{figure*}[htbp]
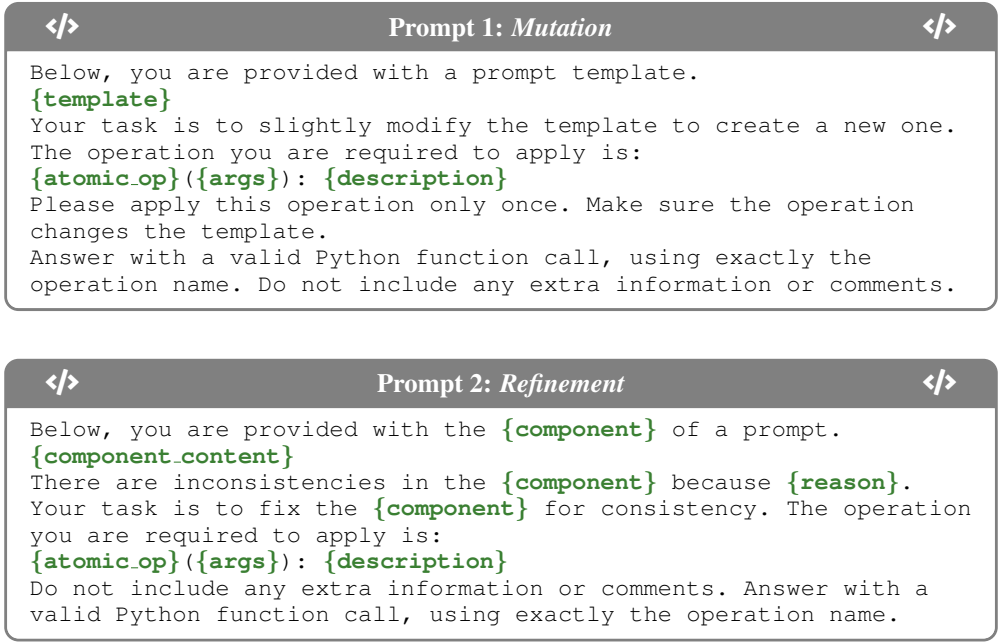

\centering
\begin{minipage}{0.8\linewidth}
\begin{metaprompt}{Mutation}{1}
Below, you are provided with a prompt template.
|\textbf{\textcolor{OliveGreen}{\{template\}}}|
Your task is to slightly modify the template to create a new one. The operation you are required to apply is:
|\textbf{\textcolor{OliveGreen}{\{atomic\_op\}}}|(|\textbf{\textcolor{OliveGreen}{\{args\}}}|): |\textbf{\textcolor{OliveGreen}{\{description\}}}|
Please apply this operation only once. Make sure the operation changes the template. 
Answer with a valid Python function call, using exactly the operation name. Do not include any extra information or comments.
\end{metaprompt}
\end{minipage}\vspace{0.6cm}\\
\begin{minipage}{0.8\linewidth}
\begin{metaprompt}{Refinement}{2}
Below, you are provided with the |\textbf{\textcolor{OliveGreen}{\{component\}}}| of a prompt.
|\textbf{\textcolor{OliveGreen}{\{component\_content\}}}|
There are inconsistencies in the |\textbf{\textcolor{OliveGreen}{\{component\}}}| because |\textbf{\textcolor{OliveGreen}{\{reason\}}}|. Your task is to fix the |\textbf{\textcolor{OliveGreen}{\{component\}}}| for consistency. The operation you are required to apply is:
|\textbf{\textcolor{OliveGreen}{\{atomic\_op\}}}|(|\textbf{\textcolor{OliveGreen}{\{args\}}}|): |\textbf{\textcolor{OliveGreen}{\{description\}}}|
Do not include any extra information or comments. Answer with a valid Python function call, using exactly the operation name. 
\end{metaprompt} 
\end{minipage}
\caption{Prompt templates for generating atomic operations, used in the mutation process}
\label{fig:metaprompts}
\end{figure*}

Recently, LLMs have been widely adopted to generate fuzzy inputs for testing software~\cite{llm_fuzzer1,llm_fuzzer2,llm_fuzzer3}, protocols~\cite{llm_fuzzer4} and firmware~\cite{llm_fuzzer5}, due to their outstanding capability of generating diverse and realistic inputs.
Inspired by these studies, we employ an LLM as a \textit{mutator} to create diverse yet semantically similar operations in an \textit{automated} manner. Specifically, we select GPT-4o~\cite{openai2024gpt4ocard} as the mutator due to its superior performance. Our mutation involves a loop process. In each iteration, a meta-template is randomly taken from the pool, mutated (Section~\ref{sec:mutation}) and refined (Section~\ref{sec:refinement}) using the LLM mutator, and then subjected to validation (Section~\ref{sec:validation}). If the mutated meta-template is valid, it is returned to the pool, otherwise it is discarded. Iterations continue until the number of valid meta-templates in the pool exceeds a predefined threshold (e.g., 100 in our study). As introduced previously, the initial element in the pool is the meta-template constructed in Section~\ref{sec:meta-template}.

\subsubsection{Mutation}
\label{sec:mutation}

To mutate a meta-template (i.e., mutate its syntax tree), the LLM-powered mutator is required to generate input arguments for an atomic operation predefined in the meta-template. Prompt~\ref{metaprompt:1} in \figref{\ref{fig:metaprompts}} shows the detailed prompt template we use to generate an atomic operation in the format of a Python-style function call
\texttt{\textbf{func}(arg1, arg2, ...)},
where \textbf{\texttt{func}} is an operation \textit{randomly} selected for fairness, and \texttt{(arg1, arg2, ...)} represent the operation parameters that the model is expected to generate. We extract the operation from the model's response, convert it to executable code, and apply the operation by executing the code.

\begin{figure}[t]
    \centering
    \includegraphics[width=0.6\linewidth]{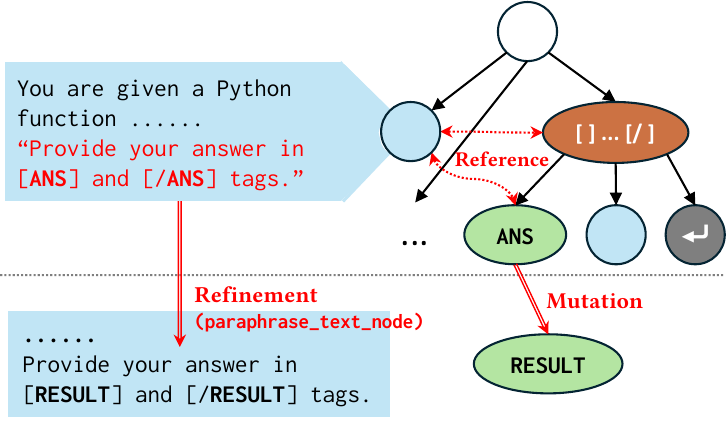}
    \caption{An example of semantic correlations between syntax tree nodes and the refinement after mutation}
    \label{fig:inconsistency}
\end{figure}

\subsubsection{Refinement}
\label{sec:refinement}

In previous sections, we design and perform atomic operations to modify syntax trees while aiming to respect the original prompt structure and semantics (i.e., \textbf{R1} and \textbf{R2}). However, this approach has an issue to address: for a syntax tree, the nodes may have semantic correlations with one another, meaning the content in one node may refer to the content in another. Yet, an atomic operation modifies only a single node and does not account for these inter-node relationships.
In such cases, a series of atomic operations may break the semantic correlations and introduce inconsistencies into the final prompt.
For example, as shown in \figref{\ref{fig:inconsistency}}, the \textcolor{Cerulean}{\textbf{\textit{blue}}} text node references both the \textcolor{Sepia}{\textbf{\textit{brown}}} format node and the \textcolor{ForestGreen}{\textbf{\textit{green}}} tag node. In the example, an atomic operation is applied to modify the tag node (i.e., \texttt{[ANS] $\rightarrow$ \texttt{[RESULT]}}). In this case, if the text node remains unchanged, these nodes would become inconsistent with each other.

To address this issue, we propose a refinement step to correct inconsistencies. Since we do not modify the structure of syntax trees (i.e., we do not add or delete nodes), we manually derive correlations between nodes and consistency-checking rules during the meta-template construction step. 
For example, in \figref{\ref{fig:inconsistency}}, the inconsistency can be detected by checking whether the ``\texttt{ANS}'' tag node or the format node has been modified.
After generating an atomic operation, we check for possible inconsistencies using its arguments. 
If inconsistencies are detected, we reuse the predefined set of atomic operations, provide detailed content for the inconsistent parts, and prompt the LLM mutator to generate corrective operations. 
Although the same set of atomic operations is used, the mutation step focuses on exploring diverse inputs, while the refinement step is dedicated to resolving inconsistencies.
Prompt~\ref{metaprompt:2} shows our detailed prompt template for refinement.
In addition, \figref{\ref{fig:inconsistency}} presents an example that addresses inconsistencies with a paraphrase operation on the text node.

\subsubsection{Validation}
\label{sec:validation}

Although the LLM-based mutator is powerful at generating creative inputs, it may produce invalid responses or operations that violate original semantics. Therefore, we validate the arguments in generated mutation or refinement operations and reject any operation that meets one of the following conditions:

\begin{enumerate}[label=\textbf{C\theenumi}]
    \item \textbf{Argument Mismatch}: The argument number or types do not match the given signature.
    \item \textbf{Semantic Inequivalence}: If the operation paraphrases a text node containing more than 10 words, we compare the similarity of sentence embedding vectors to allow for slight changes while rejecting significant semantic shifts. If the cosine similarity between the embedding vectors before and after the operation falls below a threshold (e.g., 0.85 in our study, which is considered a reasonably high threshold~\cite{similarity_thresholds}), the operation is rejected.
    \item \textbf{Description Mismatch}: The argument does not comply with the requirements outlined in the operation description.
\end{enumerate}

\section{Study Implementation}

In this section, we describe the implementation of our study, including the details of our evaluation target and experimental settings.

\subsection{Evaluated LLMs and Benchmark Tasks}
\label{sec:impl_llm_and_benchmarks}

To conduct an extensive experiment, we select various recent open-source LLMs~\cite{codellama, llama3, jiang2023mistral7b, codegemmateam2024codegemmaopencodemodels, bai2023qwentechnicalreport, hui2024qwen25codertechnicalreport, guo2024deepseekcoderlargelanguagemodel} based on the following aspects:
\begin{itemize}
    \item \textbf{Model Family}: We use popular models from the Llama, Mistral, Gemma, Qwen and DeepSeek family.
    \item \textbf{Model Type}: We use both general-purpose LLMs and code-specialized LLMs (i.e., the ones further finetuned on code datasets).
    \item \textbf{Model Size}: We use models with parameter size ranging from 6.7B to 34B.
\end{itemize}

\begin{table}[htbp]
\centering
\vspace{-0.1cm}
\caption{LLMs used in our empirical study}
\label{tab:model_list}
\begin{tabular}{|c|c|c|c|}
\hline
\textbf{MID} & \textbf{Model Name} & \textbf{Type}            & \textbf{Model Family}  \\ \hline
\textbf{M1}  & CodeLlama-7B        & \multirow{3}{*}{Code}    & \multirow{5}{*}{Llama} \\ \cline{1-2}
\textbf{M2}  & CodeLlama-13B       &                          &                        \\ \cline{1-2}
\textbf{M3}  & CodeLlama-34B       &                          &                        \\ \cline{1-3}
\textbf{M4}  & Llama3-8B           & \multirow{3}{*}{General} &                        \\ \cline{1-2}
\textbf{M5}  & Llama3.1-8B         &                          &                        \\ \cline{1-2} \cline{4-4} 
\textbf{M6}  & Mistral-7B          &                          & Mistral                \\ \hline
\textbf{M7}  & CodeGemma-7B        & \multirow{4}{*}{Code}    & Gemma                  \\ \cline{1-2} \cline{4-4} 
\textbf{M8}  & CodeQwen-7B         &                          & \multirow{2}{*}{Qwen}  \\ \cline{1-2}
\textbf{M9}  & Qwen2.5-Coder-7B    &                          &                        \\ \cline{1-2} \cline{4-4} 
\textbf{M10} & DeepSeek-Coder-6.7B &                          & DeepSeek               \\ \hline
\end{tabular}
\vspace{-0.1cm}
\end{table}

\tabref{\ref{tab:model_list}} lists the LLMs used in our study. Due to massive token consumptions and limited budget, we do not use closed-source LLMs (e.g., OpenAI models) for the complete study. We present the results of two OpenAI models in a smaller-scale study in Section~\ref{sec:disc_gpt}.

\begin{table}[h]
\centering
\caption{Benchmark tasks and metrics used in our study}
\label{tab:task_list}
\begin{tabular}{|cccc|}
\hline
\multicolumn{1}{|c|}{\textbf{TID}} & \multicolumn{1}{c|}{\textbf{Full Name}}   & \multicolumn{1}{c|}{\textbf{Task Type}}              & \textbf{Metric} \\ \hline
\multicolumn{4}{|c|}{\cellcolor[HTML]{EFEFEF}\textbf{CRUXEval}}                                                                                    \\ \hline
\multicolumn{1}{|c|}{\textbf{T1}}  & \multicolumn{1}{c|}{CRUXEval-I}      & \multicolumn{1}{c|}{\multirow{2}{1.5cm}{\centering Code Reasoning}} & \multicolumn{1}{c|}{\multirow{2}{*}{pass@5}}          \\ \cline{1-2}
\multicolumn{1}{|c|}{\textbf{T2}}  & \multicolumn{1}{c|}{CRUXEval-O}      & \multicolumn{1}{c|}{}                                & \multicolumn{1}{c|}{}          \\ \hline
\multicolumn{4}{|c|}{\cellcolor[HTML]{EFEFEF}\textbf{TestEval}}                                                                                    \\ \hline
\multicolumn{1}{|c|}{\textbf{T3}} & \multicolumn{1}{c|}{TestEval-Overall}      & \multicolumn{1}{C{1.5cm}|}{\multirow{4}{1.5cm}{\centering Test Case Generation}} &   \\ \cline{1-2} 
\multicolumn{1}{|c|}{\textbf{T4}}  & \multicolumn{1}{c|}{TestEval-Line}   & \multicolumn{1}{c|}{}                                & \multirow{2}{*}{test pass rate}  \\ \cline{1-2}
\multicolumn{1}{|c|}{\textbf{T5}}  & \multicolumn{1}{c|}{TestEval-Branch} & \multicolumn{1}{c|}{}                                &   \\ \cline{1-2} 
\multicolumn{1}{|c|}{\textbf{T6}}  & \multicolumn{1}{c|}{TestEval-Path}   & \multicolumn{1}{c|}{}                                &   \\ \hline
\multicolumn{4}{|c|}{\cellcolor[HTML]{EFEFEF}\textbf{CoderUJB}}                                                                                    \\ \hline
\multicolumn{1}{|c|}{\multirow{2}{*}{\textbf{T7}}} &
  \multicolumn{1}{c|}{\multirow{2}{*}{CoderUJB-Defect}} &
  \multicolumn{1}{c|}{Defect} &
  \multirow{2}{*}{accuracy} \\
\multicolumn{1}{|c|}{}             & \multicolumn{1}{c|}{}                & \multicolumn{1}{c|}{Detection}                       &                 \\ \hline
\multicolumn{1}{|c|}{\multirow{2}{*}{\textbf{T8}}} &
  \multicolumn{1}{c|}{\multirow{2}{*}{CoderUJB-TestGenIssue}} &
  \multicolumn{1}{c|}{Test Case} &
  \multirow{2}{*}{pass\_compile@1} \\
\multicolumn{1}{|c|}{}             & \multicolumn{1}{c|}{}                & \multicolumn{1}{c|}{Generation}                      &                 \\ \hline
\end{tabular}
\end{table}

\tabref{\ref{tab:task_list}} lists the benchmark tasks along with their corresponding evaluation metrics used in our study. Since correctness is a crucial factor for model users, a consistent set of correctness-based metrics (i.e., accuracy, pass@k, and pass rate) are used across all tasks, all of which calculate or estimate the proportion of correct generations. In specific, these tasks belong to the following three benchmarks:

\begin{itemize}
    \item \textbf{CRUXEval}~\cite{cruxeval}: CRUXEval is a benchmark designed to evaluate Python code reasoning, understanding, and execution through input and output prediction tasks. T1 and T2 challenge LLMs to predict the input and output of a Python program, respectively.
    \item \textbf{TestEval}~\cite{testeval}: TestEval is a benchmark for evaluating LLMs in Python test case generation, revealing current limitations in LLMs’ ability to understand program logic and execution paths. T3 to T6 challenge LLMs to generate test cases based on function body and function description.
    \item \textbf{CoderUJB}~\cite{coderujb}: CoderUJB is a benchmark for evaluating LLMs on Java programming tasks based on Defects4J~\cite{defects4j}, revealing their strengths and limitations in test generation and defect detection. T7 challenges LLMs to detect whether a function has defects, while T8 challenges LLMs to generate test cases based on detailed issue reports.
\end{itemize}

\subsection{Experimental Settings}

\subsubsection{Data Preparation and Statistics}

\begin{table}[htbp]
\caption{Statistics of the evaluated benchmark tasks.}
\centering
\label{tab:stats}
\begin{tabular}{@{}cccc@{}}
\toprule
\textbf{TID} & \textbf{\#instance} & \textbf{avg. \#op} & \textbf{avg. \#token} \\ \midrule
\textbf{T1}  & 800                 & 4.61                     & 277.68                \\ \dashline
\textbf{T2}  & 800                 & 5.08                     & 256.61                \\ \dashline
\textbf{T3}  & 100                 & 5.1                      & 564.81                \\ \dashline
\textbf{T4}  & 100                 & 4.2                      & 5,021                  \\ \dashline
\textbf{T5}  & 100                 & 4.03                     & 3,695                  \\ \dashline
\textbf{T6}  & 100                 & 4.25                     & 3,233                  \\ \dashline
\textbf{T7}  & 100                 & 3.7                      & 1,202                  \\ \dashline
\textbf{T8}  & 100                 & 4.72                     & 412.63                \\ \bottomrule
\end{tabular}
\vspace{-0.1cm}
\end{table}

We generate 99 mutated prompt templates for each benchmark task by executing the LLM-based mutation loop described in Section~\ref{sec:llm_mut} with a threshold of 100. Including the original prompt template, this results in a total of 100 prompt templates for each benchmark task. \tabref{\ref{tab:stats}} shows the following statistics: number of instances \textit{per task} (i.e., data samples used to fill out a prompt template), average number of atomic operations applied \textit{per template}, and average number of prompt tokens \textit{per instance} (estimated using OpenAI's \textit{tiktoken}\footnote{\url{https://github.com/openai/tiktoken}} library and \texttt{cl100k\_base} encoding). For T1 and T2, we use all the data instances in the benchmark. For other tasks, due to computational constraints (i.e., significantly longer generation and testing time compared to T1 and T2), we use the first 100 data instances in the benchmarks. Based on the statistics, the total number of input tokens for our study is approximately 1.84 billion tokens\footnote{$((277.68 + 256.61) \times 800 + (564.81 + 5021 + 3695 + 3233 + 1202 + 412.63) \times 100) \times 100 \mathrm{~templates} \times 10 \mathrm{~models}$}.

\subsubsection{Manual Verification}
\label{sec:manual_check}

As described in Section~\ref{sec:refinement} and Section~\ref{sec:validation}, we employ a validation process during mutation to filter out operations with semantic-breaking arguments. Additionally, to further enhance the trustworthiness of our study, we perform a manual check after data preparation to verify that all mutated prompt templates are semantically similar. Specifically, we inspect the arguments of all atomic operations that paraphrase the content of a node and compare them with the original version. Following this inspection, we confirm that the mutated prompt templates are indeed semantically similar, and therefore, we keep all of them for our experiments.

\subsubsection{Configuration}

\begin{table}[t]
\caption{Sampling parameters. ``temp.'' refers to temperature.}
\centering
\label{tab:gen_params}
\begin{tabular}{@{}cccc@{}}
\toprule
                  & \textbf{temp.} & \textbf{max \#token} & \textbf{\#generation}          \\ \midrule
\textbf{CRUXEval} & 0.8                  & 100                     & 10                  \\ \dashline
\textbf{TestEval} & 0                    & 256                     & 10 (T3); 1 (others) \\ \dashline
\textbf{CoderUJB} & 0.2                  & 30 (T7); 512 (T8)       & 1 (T7); 10 (others) \\ \bottomrule
\end{tabular}
\vspace{-0.1cm}
\end{table}

We run open-source LLM inference on a Linux server equipped with NVIDIA GPUs, and we deploy LLMs based on the vLLM~\cite{kwon2023efficient} framework. We use the \textit{all-mpnet-base-v2} embedding model\footnote{\url{https://huggingface.co/sentence-transformers/all-mpnet-base-v2}} in the validation step.
For each benchmark task, we reuse and adapt the original code for data loading, LLM inference, and evaluation to support different mutated prompt templates. We also follow their sampling parameters, as shown in \tabref{\ref{tab:gen_params}}, where \textit{max tokens} refers to the maximum number of tokens an LLM can generate in a single pass, and \textit{n} refers to the number of generations per sample. After generation and evaluation (i.e., \circnumwhite{1} - \circnumwhite{5} in \figref{\ref{fig:approach}}), we collect the metric values of each prompt template and conduct further analysis.

\subsubsection{RQ1}
\label{sec:rq1_setting}

In RQ1, we are interested in whether prompt mutations affect the evaluation of a single LLM. Since different benchmark tasks use different evaluation metrics, we require unitless and relative measures.
Therefore, we use Z-score, which is defined as
$$Z=(x_0-\overline{x})/\sigma.$$
In this formula, $x_0$ refers to the metric value of the original prompt template, $\overline{x}$ is the mean of the metric values across all prompt templates, and $\sigma$ is the standard deviation of the metric values:
$\overline{x} = \frac{1}{n}\sum_{i=0}^{n-1} x_i, \sigma = \sqrt{\frac{1}{n}\sum_{i=0}^{n-1}(x_i - \overline{x})^2}, n=100.$
This measure represents the distance from the mean in terms of standard deviation.

In addition, we believe that if a high improvement in performance resulting from mutating the prompt template is detected, it can also indicate substantial prompt sensitivity. Therefore, we propose a new metric named \textit{maximum performance improvement} (abbreviated as MPI), which is defined as
$$\mathrm{MPI} = \max_{i=0}^{n-1} (x_i - x_0)/x_0, n=100.$$
This measure represents the maximum improvement achieved by the mutated prompt templates compared to the original one.

\subsubsection{RQ2}
\label{sec:rq2_setting}

In RQ2, we are interested in whether prompt mutations affect the trustworthiness of the ranking results of benchmarks (i.e., evaluation of a set of LLMs). 
Therefore, we use Kendall's W (coefficient of concordance)~\cite{kendall} to measure ranking agreement among all prompt templates. 
It is defined as
$$W = 12S / m^2(n^3-n),$$
where $m$ refers to the number of \textit{judges} (i.e., 100 prompt templates in our study), $n$ refers to the number of \textit{objects} (i.e., 10 LLMs under evaluation in our study), and $S$ refers to the sum of squared deviations.

Specifically, $S$ can be calculated as follows:
$S=\sum_{i=1}^{n} (R_i - \overline{R})^2, n=10.$
$R_i$ refers to the total ranking given to object $i$ (i.e., LLM $M_i$), and $\overline{R}$ is the mean value of total rankings among all judges (i.e., prompt templates):
$R_i = \sum_{j=0}^{m-1}r_{i,j}, \overline{R} = \frac{1}{n}\sum_{i=1}^{n}R_i, n=10, m=100,$
where $r_{i,j}$ is an integer value between 1 and $n$, representing the ranking assigned to an LLM. 

The value of $W$ ranges from 0 to 1. Values closer to 1 indicate more agreement (i.e., more consistency), while values closer to 0 indicate more disagreement. We follow a previous study~\cite{mizrahi-etal-2024-state} and set the thresholds to (1) $W \ge 0.85$: strong agreement; (2) $W < 0.85$: weak to moderate agreement. 

In addition, we investigate whether prompt templates with high performance on one model can be generalized to other models. Thus, for each benchmark task, we calculate the Intersection-over-Union (IoU) for each pair of models, which is defined as
$$\mathrm{IoU} = |S_i \cap S_j| / |S_i \cup S_j|, 1 \le i < j \le 10,$$
where $S_i$ is the set of prompt templates which produce top-$k$ performance on LLM $M_i$. A high IoU value of $M_i$ and $M_j$ indicates that the top-performing prompt templates for $M_i$ are also suitable for $M_j$, and vice versa. 

\section{Study Results}
\label{sec:results}

In this section, we present the experimental results and provide findings regarding the research questions introduced in Section~\ref{sec:intro}.

\subsection{RQ1: Impact on Absolute Performance}

\subsubsection{Performance Variations} 
\label{sec:rq1_quant}

\begin{figure*}[htbp]
\centering
\begin{subfigure}{0.5\linewidth}
\centering
\includegraphics[width=\linewidth]{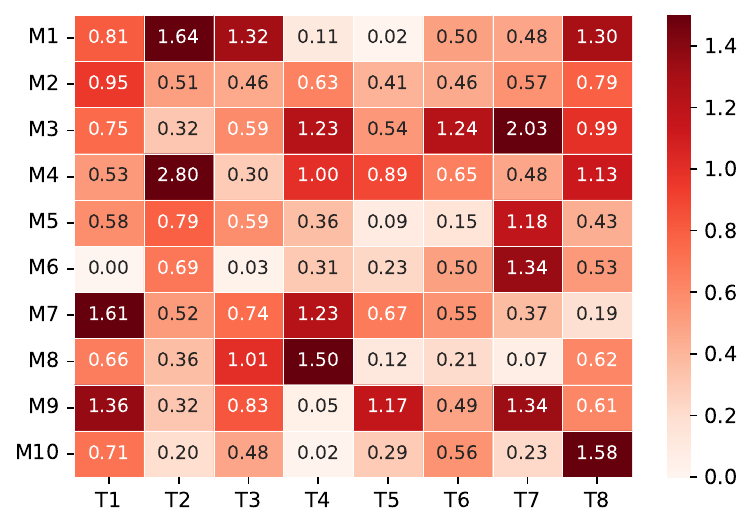}
\caption{Heatmap of absolute Z-score of model-task pairs}
\label{fig:rq1_1}
\end{subfigure}%
\begin{subfigure}{0.5\linewidth}
\centering
\includegraphics[width=\linewidth]{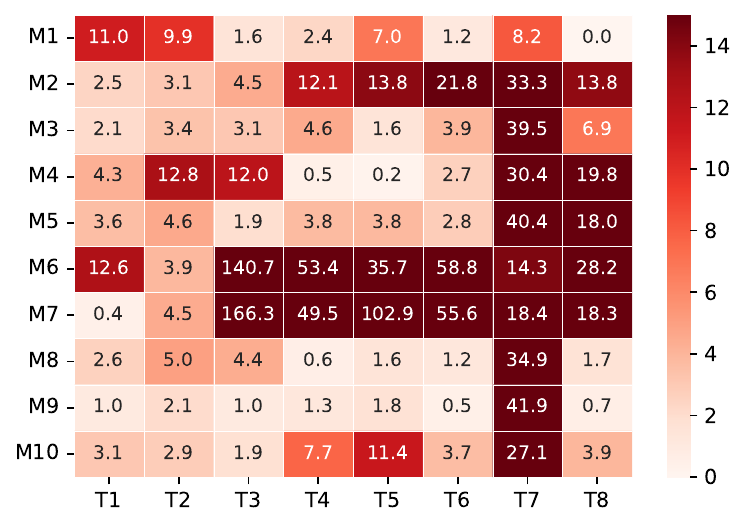}
\caption{Heatmap of MPI (\%) of model-task pairs}
\label{fig:rq1_2}
\end{subfigure}
\caption{Quantitative Results of RQ1. Darker cells indicate a larger magnitude of prompt sensitivity.}
\label{fig:rq1}
\end{figure*}

We use performance variations to quantitatively assess the impact of prompt sensitivity on the performance of an individual model. As described in Section~\ref{sec:rq1_setting}, we calculate the Z-score and the maximum performance improvement (MPI) for all model-task pairs, with the results presented as heatmaps in \figref{\ref{fig:rq1}},  and a darker cell in the heatmap indicates a higher value.

\noindent\textbf{Z-score.}
\figref{\ref{fig:rq1_1}} shows the absolute values of the Z-scores, which represent the distance between the performance with the original prompt template and the mean performance across all prompt templates. Following a previous study~\cite{mizrahi-etal-2024-state}, we consider a performance variation substantial if the Z-score exceeds 1. Based on this criterion, 24\% of the model-task pairs in our study exhibit substantial variations. Furthermore, the phenomenon of prompt sensitivity is widespread, with variations observed across multiple models for all tasks, and across multiple tasks for all models.

\noindent\textbf{Max Performance Improvement.}
\figref{\ref{fig:rq1_2}} illustrates the maximum performance improvement (MPI) in percentages relative to the original prompt template. Similar to the Z-score, substantial MPI is also prevalent. In the heatmap, 39\% of the model-task pairs show the possibility of improvement exceeding 10\%. Although various prompt engineering techniques (e.g., chain-of-thought prompting, retrieval-augmented generation, and iterative refinements) have been developed to enhance model performance on specific code benchmarks, our study demonstrates that minor semantically similar prompt template mutations can also yield substantial performance gains for a given model.

\finding{The performance of an individual model on code benchmarks is susceptible to minor semantically similar prompt template mutations. Thus, relying on a single arbitrary prompt template may not provide a robust evaluation of an LLM’s true capabilities and knowledge boundaries.}

\subsubsection{Qualitative Analysis}

\begin{table}[htbp]
\caption{Representative cases of prompt template mutations with small text edit distance ($\le$ 2 operations) but large performance difference. ``\texttt{\{\}}'' represents a text placeholder, and ``\texttt{<TAG>}'' represents the content of a tag node.}
\centering
\label{tab:rq1_case}
\begin{tabular}{|c|c|>{\ttfamily\small}c|c|}
\hline
\textbf{TID}        & \textbf{MID} & \normalfont\normalsize\textbf{Effect of Mutation}              & \textbf{Perf. Diff.}    \\ \hline
\multirow{2}{*}{T3} &
  M6 &
  \multirow{2}{*}{\begin{tabular}[c]{@{}c@{}}Program under test:\\ $\downarrow$ \\ PROGRAM UNDER TEST:\end{tabular}} &
  \begin{tabular}[c]{@{}c@{}}0.32 $\rightarrow$ 0.39\\ (\textcolor{red}{+23\%})\end{tabular} \\ \cline{2-2} \cline{4-4} 
 &
  M7 &
   &
  \begin{tabular}[c]{@{}c@{}}0.25 $\rightarrow$ 0.38\\ (\textcolor{red}{+50\%})\end{tabular} \\ \hline
T4 &
  M7 &
  \begin{tabular}[c]{@{}c@{}}\small Function description for \{\}:\\ $\downarrow$\\ \small Explanation of the function \{\}:\end{tabular} &
  \begin{tabular}[c]{@{}c@{}}0.58 $\rightarrow$ 0.51\\ (\textcolor{ForestGreen}{-12\%})\end{tabular} \\ \hline
\multirow{2}{*}{T6} &
  M2 &
  \multirow{2}{*}{\begin{tabular}[c]{@{}c@{}}\footnotesize Your test method should begin with: \\ $\downarrow$\\ \footnotesize Initiate your test method with:\end{tabular}} &
  \begin{tabular}[c]{@{}c@{}}0.78 $\rightarrow$ 0.46\\ (\textcolor{ForestGreen}{-40\%})\end{tabular} \\ \cline{2-2} \cline{4-4} 
 &
  M6 &
   &
  \begin{tabular}[c]{@{}c@{}}0.50 $\rightarrow$ 0.55\\ (\textcolor{red}{+10\%})\end{tabular} \\ \hline
\multirow{3}{*}{T7} &
  \multirow{2}{*}{M3} &
  \begin{tabular}[c]{@{}c@{}}Question $\longrightarrow$ question\\ Answer $\longrightarrow$ answer\end{tabular} &
  \begin{tabular}[c]{@{}c@{}}0.43 $\rightarrow$ 0.55\\ (\textcolor{red}{+28\%})\end{tabular} \\ \cline{3-4} 
 &
   &
  \{<TAG>\}:  $\longrightarrow$ ¿¡!\{<TAG>\}¡¿? &
  \begin{tabular}[c]{@{}c@{}}0.43 $\rightarrow$ 0.55\\ (\textcolor{red}{+28\%})\end{tabular} \\ \cline{2-4} 
 &
  M5 &
  \multirow{2}{*}{\begin{tabular}[c]{@{}c@{}}A. $\longrightarrow$ Alpha. \\ \\ B. $\longrightarrow$ Beta.\end{tabular}} &
  \begin{tabular}[c]{@{}c@{}}0.47 $\rightarrow$ 0.58\\ (\textcolor{red}{+23\%})\end{tabular} \\ \cline{2-2} \cline{4-4}
  &
  M7 &
  &
  \begin{tabular}[c]{@{}c@{}}0.49 $\rightarrow$ 0.42\\ (\textcolor{ForestGreen}{-14\%})\end{tabular} \\ \hline
T8 &
  M6 &
  ISSUE-ID $\longrightarrow$ ISSUE-id &
  \begin{tabular}[c]{@{}c@{}}0.16 $\rightarrow$ 0.18\\ (\textcolor{red}{+10\%})\end{tabular} \\ \hline 
\end{tabular}
\end{table}

\tabref{\ref{tab:rq1_case}} presents several representative cases of minor prompt template mutations that lead to substantial shifts in model performance. Despite the small edit distances between the original and mutated templates, with prompt semantics remaining similar, we observe absolute performance differences ranging from 10\% to 50\% in \tabref{\ref{tab:rq1_case}}. 

Furthermore, in \tabref{\ref{tab:rq1_case}}, we find that the performance impact of similar operations may not be consistent, depending on the model and the task. For instance, in cases (T3, M6), (T3, M7), (T7, M3), and (T8, M6), modifying the cases of only a few words results in substantial performance increase. However, two of them change the lower cases to upper cases, while the other two perform opposite operations. In addition, in cases (T6, M2) and (T6, M6), text paraphrasing leads to a 10\% performance increase for M6, but a 40\% decrease for M2. Even within the same task, a mutation could lead to opposite performance changes for different models. Thus, it is challenging to predict a pattern that ensures performance improvement. A similar phenomenon can also be observed in cases (T7, M5) and (T7, M7). 

These examples support Finding~\ref{1} in Section~\ref{sec:rq1_quant}. They again underscore the impact of prompt template choices, demonstrating that having only one prompt template might be insufficient for robust model evaluations.

\subsubsection{Correlation between Sensitivity and Capability}

\begin{table}[htbp]
\caption{Pearson correlation coefficients between Z-scores and mean performance metrics for all benchmark tasks}
\centering
\label{tab:rq1_corr}
\begin{tabular}{@{}ccccccccc@{}}
\toprule
    & \textbf{T1} & \textbf{T2} & \textbf{T3} & \textbf{T4} & \textbf{T5} & \textbf{T6} & \textbf{T7} & \textbf{T8} \\ \midrule
$r$ & 0.71         & -0.40        & 0.44         & 0.01         & 0.06         & 0.02         & 0.32         &  -0.26           \\ \bottomrule
\end{tabular}
\end{table}

To examine the relationship between prompt sensitivity and model capability for a specific task, we compute Pearson correlation coefficients~\cite{freedman2007statistics} (i.e., Pearson's $r$) between the Z-scores and mean performance across models. Performance is averaged over all prompt templates to estimate a model’s capability on a task. As shown in \tabref{\ref{tab:rq1_corr}}, the results reveal various types of correlations. Tasks T4, T5, and T6 exhibit near-zero $r$ value, indicating negligible correlations. For tasks with non-zero $r$ values, T1 shows a strong correlation, while others show weak correlations. Notably, three tasks show positive correlations, and two show negative correlations.

\finding{Reliably predicting prompt sensitivity for a task based solely on model capabilities is challenging.}

\begin{table}[ht]
\caption{Kendall's W for each benchmark task}
\centering
\label{tab:rq2}
\begin{tabular}{@{}cllllllll@{}}
\toprule
\multicolumn{1}{l}{} &
  \multicolumn{1}{c}{\textbf{T1}} &
  \multicolumn{1}{c}{\textbf{T2}} &
  \multicolumn{1}{c}{\textbf{T3}} &
  \multicolumn{1}{c}{\textbf{T4}} &
  \multicolumn{1}{c}{\textbf{T5}} &
  \multicolumn{1}{c}{\textbf{T6}} &
  \multicolumn{1}{c}{\textbf{T7}} &
  \multicolumn{1}{c}{\textbf{T8}} \\ \midrule
\textbf{All Models} &
  \multicolumn{1}{r}{.92} &
  \multicolumn{1}{r}{.90} &
  \multicolumn{1}{r}{.91} &
  \multicolumn{1}{r}{.83} &
  \multicolumn{1}{r}{.86} &
  \multicolumn{1}{r}{.92} &
  \multicolumn{1}{r}{.30} &
  \multicolumn{1}{r}{.94} \\ \dashline
\textbf{CodeLlama} &
  \multicolumn{1}{r}{.83} &
  \multicolumn{1}{r}{.92} &
  \multicolumn{1}{r}{.61} &
  \multicolumn{1}{r}{.87} &
  \multicolumn{1}{r}{.89} &
  \multicolumn{1}{r}{.91} &
  \multicolumn{1}{r}{.01} &
  \multicolumn{1}{r}{.49} \\ \dashline
\multicolumn{1}{l}{\textbf{Llama 3.x}} &
  .92 &
  1.0 &
  .81 &
  .49 &
  .74 &
  .88 &
  .49 &
  .06 \\ \dashline
\textbf{Llama (All)} &
  .91 &
  .94 &
  .68 &
  .59 &
  .76 &
  .82 &
  .15 &
  .85 \\ \dashline
\textbf{Qwen} &
  1.0 &
  1.0 &
  .96 &
  .88 &
  1.0 &
  1.0 &
  .23 &
  .71 \\ \bottomrule
\end{tabular}
\end{table}

\subsection{RQ2: Impact on Relative Performance}

\subsubsection{Kendall's W}

In addition to the absolute performance differences introduced by prompt template mutations (RQ1), the relative performance differences among models are also important.
As described in Section~\ref{sec:rq2_setting}, we use Kendall's W to measure the agreement (i.e., consistency) of the rankings across all prompt templates, as shown in \tabref{\ref{tab:rq2}}, which presents the results for all models as well as for specific model families.
For rankings across all models, despite substantial performance variations in individual models (RQ1), most tasks show moderate to strong ranking agreement. This result is likely due to the significant performance gap between models of different architectures or generations. As an exception, task T7 exhibits poor ranking agreement, suggesting that this benchmark task fails to provide consistent rankings and therefore lacks robustness.
However, when considering the ranking within a specific model family, stronger disagreements become more apparent.
Within the CodeLlama family, the 7B, 13B, and 34B models, which share similar architectures and are trained and fine-tuned on the same dataset, are expected to exhibit consistent ranking orders (i.e., larger variants outperform smaller ones). However, the Kendall's W values for T3, T7, and T8 drop significantly, indicating a substantial increase in ranking disagreement (i.e., smaller variants outperform larger ones). Similarly, within the Llama 3.x and Qwen model series, we can also observe an increase in disagreement for some tasks. 

\finding{Model rankings from specific code benchmarks are prone to minor semantically similar prompt template mutations, especially within a particular model family. Thus, relying on a single arbitrary prompt template may fail to provide robust capability rankings across models.}

\subsubsection{IoU}

\begin{table}[htbp]
\caption{Mean IoU for each benchmark task, averaged over all non-repetitive model-model pairs}
\centering
\label{tab:rq2_iou}
\begin{tabular}{@{}ccccccccc@{}}
\toprule
 & \textbf{T1} & \textbf{T2} & \textbf{T3} & \textbf{T4} & \textbf{T5} & \textbf{T6} & \textbf{T7} & \textbf{T8} \\ \midrule
$k=1$  & .04 & 0   & 0   & .02 & .04 & .04 & 0   & 0 \\ \dashline
$k=5$  & .05 & .05 & .02 & .04 & .11 & .04 & .03 & .05 \\ \dashline
$k=10$ & .10 & .08 & .05 & .06 & .12 & .06 & .05 & .11 \\ \dashline
$k=20$ & .16 & .16 & .11 & .13 & .16 & .13 & .11 & .21 \\ \bottomrule
\end{tabular}
\end{table}

Furthermore, we use the Intersection-over-Union (IoU) metric to assess the degree of overlap between prompt templates that yield top-$k$ performance for different model pairs, thereby evaluating the transferability of high-quality prompt templates across models.
\tabref{\ref{tab:rq2_iou}} presents the mean IoU values (averaged over all model-model pairs) across eight benchmark tasks for different values of $k$, where $k$ is set to $[1,5,10,20]$. Overall, the results reveal that the IoU values are consistently low, despite the slight increase with higher $k$ value. Especially for $k \le 10$, the IoU values are mostly close to 0, suggesting poor overlap in the top-performing templates across models.

\finding{Transferring top-performing prompt templates between models is challenging, leading to unfair comparisons and unreliable rankings in current code benchmarks that rely on a single arbitrary prompt
template.}

\section{Discussion}

In this section, we first discuss the results of our case studies, which examine the impact of randomness and the performance of some other models. Finally, we address the threats to validity.

\subsection{Impact of Randomness}
\label{sec:disc_temp}

Various sampling parameters, such as temperature, control the randomness of generation~\cite{humaneval}. To completely avoid the randomness introduced in the decoding stage, a common approach is to apply the greedy decoding strategy. However, in many software engineering benchmarks (e.g., two benchmarks used in our study, CRUXEval~\cite{cruxeval} and CoderUJB~\cite{coderujb}), due to the difficulty of tasks, researchers tend to utilize randomness to explore the vast output space and find more correct answers. Therefore, both randomness and prompt template mutation influence the final metric values. 

We conduct this case study to discuss (1) whether the impact of prompt template mutation is statistically significant, even when there is relatively high randomness in decoding; and (2) whether there is interaction between prompt template mutation and randomness. For simplicity, we only adjust the temperature parameter to control randomness, leaving the other sampling parameters unchanged. Task T2 (i.e., CRUXEval-O) is used for this case study. Specifically, we set the temperatures to $[0.2, 0.4, 0.6, 0.8, 1.0]$, and repeat five times for each model-temperature pair. Then we conduct two-way analysis of variance (ANOVA)~\cite{girden1992anova}, which examines the influence of prompt template and temperature on the performance and assesses if there is any interaction between the two factors.

\begin{table}[htbp]
\centering
\caption{F values of prompt template, temperature and their interaction in two-way analysis of variance for all models (p $<$ 1e-5 for all values in the table)}
\label{tab:dicussion_temp}
\begin{tabular}{@{}lrrrrr@{}}
\toprule
                         & \textbf{M1} & \textbf{M2} & \textbf{M3} & \textbf{M4} & \textbf{M5}   \\ \midrule
\textbf{Prompt Template} & 78          & 56          & 68          & 163         & 124           \\ \dashline
\textbf{Temperature}     & 19,419      & 21,988      & 17,957      & 26,068      & 21,036        \\ \dashline
\textbf{Interaction}     & 5           & 2           & 5           & 2           & 3             \\ \midrule
                         & \textbf{M6} & \textbf{M7} & \textbf{M8} & \textbf{M9} & \textbf{M10}  \\ \midrule
\textbf{Prompt Template} & 551         & 422         & 6,888       & 92          & 4325          \\ \dashline
\textbf{Temperature}     & 19,374      & 3,238       & 25,755      & 10,559      & 26,664        \\ \dashline
\textbf{Interaction}     & 4           & 5           & 60          & 3           & 3             \\ \bottomrule
\end{tabular}
\end{table}

\tabref{\ref{tab:dicussion_temp}} shows the test statistics (i.e., F-values) of the analysis. In the analysis, the null hypothesis ($H_0$) is that there is no difference between group means, and we reject the null hypothesis according to the $p$ values. We derive the following observations:

\circnum{1} Prompt template mutation has \textbf{statistically significant} influence on performance, even when temperature is high. 

\circnum{2} A larger F-value indicates a larger difference between the group means. Therefore, temperature has a higher influence than our prompt template mutation when both factors are changing.

\circnum{3} Although the p-value indicates an interaction between prompt template mutation and temperature, the magnitude of this interaction is much smaller than the individual effects of each factor. Therefore, we can disregard it and conclude that the impacts of prompt template mutation and temperature are independent.

\subsection{Results of Other Models}
\label{sec:disc_gpt}

In addition to the LLMs studied in previous sections, proprietary models (e.g., OpenAI models) are widely used. Recently, reasoning models (e.g., OpenAI o3-mini~\cite{openai_o3_mini} and DeepSeek-R1~\cite{deepseekr1}) have also gained significant attention due to their state-of-the-art performance in complex tasks. 
Due to high token consumption and budget constraints, we conduct a smaller-scale study using the following models: GPT-3.5-Turbo, GPT-4o-mini and DeepSeek-R1-Distill-Qwen-32B. For OpenAI models, the sampling parameters match those listed in \tabref{\ref{tab:gen_params}}. For the DeepSeek-R1-Distill model, we set the temperature to 0.6 to prevent incoherent outputs, as strongly recommended in the official documentation~\cite{huggingface_deepseek}. We limit the maximum length of reasoning tokens to 2,048 by implementing a logits processor. In addition, when using the R1 model for CRUXEval tasks (i.e., T1 and T2), the maximum length of completion tokens is increased to 300 to avoid truncating the answers.

Specifically, the experiments in this case study are downsized to 10 prompt templates (1 original and 9 mutated) while maintaining the number of instances per task. To improve the representativeness of mutations, we develop a set of diversity-based sampling rules for prompt template selection:

\begin{enumerate}
    \item Mutated prompt templates are grouped based on the number of distinct operation types, and at least one template is selected from each group.
    \item Within each group, templates are sorted by the percentage increase or decrease in model performance compared to the original template (averaged across all open-source models). Templates with the most significant performance differences are selected until the required number of templates is reached.
\end{enumerate}

\begin{table}[ht]
\centering
\caption{Results of selected models. G-3.5: GPT-3.5-Turbo, G-4: GPT-4o-mini, R1-32B: DeepSeek-R1-Distill-Qwen-32B.}
\label{tab:gpt_results}
\begin{tabular}{@{}cccccccccc@{}}
\toprule
 & \textbf{} & \textbf{T1} & \textbf{T2} & \textbf{T3} & \textbf{T4} & \textbf{T5} & \textbf{T6} & \textbf{T7} & \textbf{T8} \\ \midrule
\multirow{3}{*}{\textbf{Z-score}}  & \textbf{G-3.5} & 0.48 & 0.89 & -1.06& 0.31 & 0.02  & -0.86& -2.13 & -0.41\\ \pdashline{2-10}
                                   & \textbf{G-4}   & 1.39 & 0    & 0.17 & 0.85 & -1.33 & 0    & -2.17 & 0.37 \\ \pdashline{2-10}
                                   & \textbf{R1-32B}& 0.40 & 0.67 & 1.74 & 0.40 & 0.64  & -0.74& -1.02 & -2.45\\ \midrule
\multirow{3}{*}{\textbf{MPI (\%)}} & \textbf{G-3.5} & 1.42 & 0.79 & 2.29 & 0.96 & 1.54  & 1.24 & 32.6  & 52.7 \\ \pdashline{2-10}
                                   & \textbf{G-4}   & 0    & 1.20 & 0.71 & 0.16 & 0.43  & 0.25 & 20.0  & 8.33 \\ \pdashline{2-10}
                                   & \textbf{R1-32B}& 0.41 & 0.42 & 0    & 1.12 & 0.44  & 4.17 & 19.6  & 116  \\ \midrule
\multicolumn{2}{c}{\textbf{Kendall's W}}            & 1.00 & 0.75 & 0.49 & 0.75 & 0.76  & 1.00 & 0.21  & 0.91 \\ \bottomrule
\end{tabular}%
\end{table}

\tabref{\ref{tab:gpt_results}} presents the results of the selected proprietary and reasoning models, which are similar to those of the open-source models discussed in Section~\ref{sec:results}. The Z-scores (33\% of cases identified as exhibiting \textit{substantial performance variations}) indicate that the selected models are also sensitive to semantically similar prompt template mutations (\textbf{Finding}~\ref{1}). From another perspective, substantial performance improvements are observed only for tasks T7 and T8, while for other tasks, the mutated prompt templates do not yield significant gains. This finding suggests the limited transferability of top-performing prompt templates across the selected models (\textbf{Finding}~\ref{4}).
In addition, the values of Kendall's W (with 5 out of 8 cases identified as exhibiting \textit{weak to moderate agreement}) suggest that prompt sensitivity introduces inconsistencies in model rankings (\textbf{Finding}~\ref{3}). Based on these results, we conclude that current code benchmarks also encounter prompt sensitivity issues when evaluating proprietary and reasoning models.

\subsection{Threats to Validity}

\subsubsection{Internal Threats} For internal validity, we focus on the \textit{trustworthiness} of our study.

\noindent\textbf{Implementation Issue.} Postprocessing refers to extracting the exact answer (e.g., function body) from a model's response. This is a common practice in LLM evaluation, since LLMs often generate extraneous content (e.g., explanations and comments), which causes the evaluation code to fail and further leads to untrustworthy conclusions. As described in Section~\ref{sec:impl_llm_and_benchmarks}, we reuse and adapt the benchmark code for generation and evaluation. However, the adapted postprocessing code may not work for new models or new mutated prompt templates. This issue is inherited from prior work on benchmarks. 
To address it in our study, we manually sample and inspect the generation results after postprocessing. During this inspection, we identified only a few unhandled edge cases, which were subsequently addressed. 
After fixing the implementation, we re-run the postprocessing. 
In rare cases where a model generates incorrect tag node tokens or unrelated tokens, causing postprocessing to fail, we do not manually correct the generations, as we attribute the issue to the model’s suboptimal instruction-following capability. 
With the above efforts, we believe that our study provides a fair evaluation and comparison.

\noindent\textbf{Randomness in Generation.} Along with our prompt template mutation, randomness is also a influential factor on model response, which may affect our findings. We mitigate this issue with the following efforts. Firstly, for the TestEval and CoderUJB benchmarks, we follow their original experimental settings and use very low temperatures (i.e., 0 and 0.2). Since other sampling parameters are not enabled, this makes the model response deterministic (for zero temperature) or nearly deterministic (for 0.2 temperature). Secondly, for the CRUXEval benchmark which uses a higher temperature, we conduct replication experiments, and the analysis results in Section~\ref{sec:disc_temp} show that our prompt template mutation has statistically significant effect even with a relatively high temperature (e.g., 0.8 and 1.0). Therefore, we believe that our findings are reliable in current settings.

\subsubsection{External Threats} For external validity, we focus on the \textit{generalizability} of our study. The results of our study are based on the selected LLMs and the benchmark tasks, and there is a potential generalizability issue. To mitigate this issue, we select various representative open source models, considering many criteria, including model family, popularity, scale, and use cases. In addition, we conduct a case study in Section~\ref{sec:disc_gpt}. In particular, we select GPT-3.5-Turbo and GPT-4o-mini as representative proprietary models, and DeepSeek-R1-Distill-Qwen-32B as a representative reasoning model. For the benchmark tasks, we select the ones from recent state-of-the-art researches, which emcompass code reasoning, defect detection and test code generation. 
Some other code benchmarks~\cite{humaneval,mbpp} are excluded from the scope of our study due to their relatively simple prompts, which offer limited opportunities for mutation.

\section{Conclusion and Future Work}

In this paper, we present an empirical study that investigates the impact of prompt sensitivity on code benchmarks, demonstrating how slight, semantically similar modifications to prompt templates can lead to significant variations in both absolute and relative model performance. Through extensive experiments across multiple tasks and models, we find that prompt sensitivity is a common phenomenon in code benchmarks, potentially introducing inconsistencies in performance rankings and unreliable performance estimations. These findings underscore the need to enhance the robustness of evaluation in future code benchmarks.
While our study offers valuable insights, it leaves several areas unexplored. Specifically, we focus on benchmark tasks that evaluate an LLM in isolation. Future research could explore more complex scenarios, such as tasks involving multi-LLM interactions.

\bibliographystyle{unsrt}
\bibliography{ref}

\end{document}